\documentclass[journal,twoside]{IEEEtran}
\usepackage{cite}
\usepackage{amsmath,amssymb,amsfonts}
\usepackage{algorithmic}
\usepackage{graphicx}
\usepackage{textcomp}
\usepackage{wrapfig}

\usepackage[hidelinks]{hyperref}

\usepackage[per-mode=symbol]{siunitx}
\DeclareSIUnit[]{\pixel}{pixel}

\usepackage[acronym]{glossaries}
\newacronym{dac}{DAC}{digital-to-analog converter}
\newacronym{adc}{ADC}{analog-to-digital converter}
\newacronym{mcu}{MCU}{microcontroller unit}
\newacronym{dsp}{DSP}{digital signal processing}
\newacronym{usb}{USB}{Universal Serial Bus}
\newacronym{snr}{SNR}{signal-to-noise ratio}
\newacronym{pid}{PID}{proportional-integral-derivative controller}
\newacronym{pd}{PD}{photodiode}
\newacronym{ld}{LD}{laser diode}
\newacronym{lp}{LP}{low-pass}
\newacronym{sp}{SP}{short-pass}
\newacronym{bp}{BP}{band-pass}
\newacronym{daq}{DAQ}{data acquisition system}
\newacronym{bs}{BS}{beam splitter}
\newacronym{sod}{SOD}{standoff distance}
\newacronym{lmd}{LMD}{laser-metal deposition}
\newacronym{slm}{SLM}{selective laser melting}
\newacronym{lpbf}{LPBF}{laser powder bed fusion}
\newacronym{roi}{ROI}{region of interest}
\newacronym{na}{NA}{numerical aperture}
\newacronym{fov}{FOV}{field of view}
\newacronym{tia}{TIA}{transimpedance amplifier}
\newacronym{pc}{PC}{computer}
\newacronym{fwhm}{FWHM}{full width at half maximum}
\newacronym{cw}{CW}{continuous wave}
\newacronym{psd}{PSD}{position sensitive detector}
\newacronym{oct}{OCT}{optical coherence tomography}
\newacronym{fpga}{FPGA}{field-programmable gate array}
\newacronym{lidar}{LiDAR}{light detection and ranging}

\begin{document}
\title{Hybrid optical sensor for combined thermal and dimensional monitoring in laser processing}

\author{Simone~Donadello \thanks{The author was with the Department of Mechanical Engineering, Politecnico di Milano, 20156 Milan, Italy. He is now with the Quantum Metrology and Nanotechnologies Division, Istituto Nazionale di Ricerca Metrologica, INRIM, 10135 Turin, Italy.\\
Published version DOI: \href{https://doi.org/10.1109/JSEN.2024.3477278}{10.1109/JSEN.2024.3477278}}}

\maketitle

\begin{center}
	\includegraphics[width=\columnwidth]{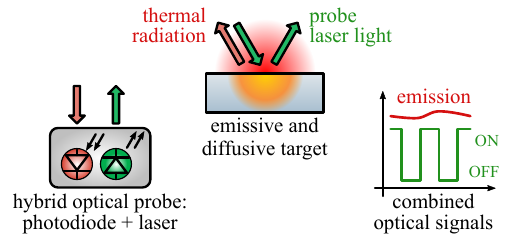}%
\end{center}

\begin{abstract}
	Optical measurements enable non-contact and high-speed monitoring of physical processes, offering a non-invasive and versatile approach across a wide range of fields, from scientific research to industrial applications. This work presents an optical sensor capable of simultaneously measuring both the distance and thermal emission from surfaces, based on a simple laser diode probe. The principle of operation integrates triangulation with pyrometry in a single device, leveraging the monitor photodiode embedded within the probe module. By alternating the probe laser emission, the system can rapidly switch between dimensional and thermal measurements, resulting in combined data acquisition. The proposed method is compact, easy to integrate, and cost-effective. The hybrid sensor is demonstrated in a laser processing setup, where a metallic target is heated and melted by a high-power laser beam. Its inline operation allows for real-time dynamic measurements of melt pool distance and radiance, in a coaxial and self-aligning configuration. This innovative approach can be applied to various fields, such as remote environmental sensing and closed-loop control systems for stabilizing high-temperature processes, including laser welding and additive manufacturing.
\end{abstract}

	
	\section{Introduction}
	
	\IEEEPARstart{O}{ptical} monitoring techniques have revolutionized our ability to observe and study physical processes, from materials science and engineering, to environmental monitoring and biomedical research. Optical probes offer non-invasive and versatile means of measurement, even in challenging contexts. Their utility is particularly evident in laser material processing, where a precise control over thermal and dimensional parameters is required. Laser-based sensors and imaging systems introduce important advantages in this field, such as high-speed detection, operation without physical contact, and compatibility with high-temperature environments, making them ideal for in-process monitoring applications \cite{evertonReview2016,youReview2014}. These techniques enable deeper comprehension of the underlying physical processes in several laser manufacturing applications. Optical diagnostic tools are ideal for characterizing the complex nature and fast dynamics of laser-induced plasmas \cite{harilalOptical2022}, including emission spectroscopy \cite{aragonCharacterization2008}, holography \cite{pangovskiHolographic2016}, and self-mixing interferometry \cite{donadelloTimeresolved2020}. In additive manufacturing processes, like laser cladding, \gls{lmd}, or \gls{slm}, where heat accumulation effects are crucial, photodiodes and thermal cameras are often used to measure the melt pool temperature and geometry \cite{heInsitu2019,mccannInsitu2021}. Optical triangulation can be exploited to monitor the deposition height, giving insights on the process efficiency and stability \cite{donadelloInterplay2022}. Closed-loop control systems utilizing pyrometers enable real-time stabilization of the molten metal temperature \cite{renkenInprocess2019}. \Gls{oct} has garnered great interest for accurate inline dimensional control in laser welding and ablation \cite{websterAutomatic2014}. Digital imaging techniques can be implemented for process quality inspection in laser cutting \cite{pacherRealtime2020}. Nonetheless, a multi-sensor approach is often necessary to gain a comprehensive insight into the physical process \cite{maffiaCoaxial2023,gutknechtMutual2021}, monitoring different parameters together with data fusion and machine learning methods~\cite{kongMultisensor2020}.
	
	This work presents the design, implementation, and validation of an innovative technique for combined dimensional and thermal monitoring utilizing a single instrument. The optical sensor, based on a simple laser probe and its monitoring photodiode, integrates triangulation with pyrometry \cite{donadelloCombined2023}. The thermal measurement leverages the photodiode embedded within the laser diode module, typically used to monitor the laser emission power. A complementary study demonstrated the ability of such a photodiode to measure the internal package temperature \cite{foldesyTemperature2022}. Here, the thermal measurement is remote, with the external radiation being collected over a field of view defined by the triangulation probe beam, exploiting the reversibility principle of light. In particular, when the probe laser is not emitting, the photodiode detects light propagating along the reversed optical path \cite{giulianiLaser2002}. By continuously switching the probe laser operation at high speed, the sensor enables self-aligned and synchronized measurements of distance and radiance of high-temperature surfaces.
	
	As a showcase of the technique, the sensor was integrated into an industrial laser processing system, successfully monitoring the standoff distance and thermal emission of a melt pool generated by a high-power laser. The coaxial measurements enabled the detection of transitions between different process regimes during the melting of stainless steel targets, identifying instabilities in the melt pool formation. While the determination of absolute temperature would require pyrometric calibrations, these are unnecessary for detecting relative deviations from operational setpoints. Indeed, the experimental validation of the sensor capability in monitoring the relative temperature and dynamics of molten metals demonstrates its realistic potential for application in various laser processes, including laser welding, cutting, and additive manufacturing. The simultaneous measurement of dimensional and thermal properties eliminates the need for additional components or complex multi-sensor setups based on separate instruments. The compactness, low-cost, and ease of implementation of this solution offer several advantages, paving the way for combined process stabilization strategies in industrial closed-loop control systems. 
	
	The underlying principle can potentially be integrated into other optical techniques based on two-way sensing, such as interferometry \cite{donadelloProbing2018}, reflectometry \cite{leal-juniorOptoElectronic2024}, fiber sensors \cite{gomesLaserInduced2024}, or into processing laser sources. This versatility allows for broader applicability in interdisciplinary measurement scenarios beyond laser material processing. Among other applications, it could be employed to inspect thin film deposition in semiconductor manufacturing \cite{hermanOptical1996}, furnaces \cite{paunaOptical2020}, plasma fluorescence \cite{martiniReactivity2017}, or high-temperature combustion processes \cite{fuTemperature2014}. Additionally, the detection method may be used in environmental monitoring, ranging from the study of climate change by measuring the brightness and reflectivity of glaciers and snow deposits \cite{konigMeasuring2001} to the surveillance of volcanic magmas \cite{spampinatoVolcano2011}. Further miniaturization and sensitivity enhancements could open intriguing possibilities for developing novel integrated sensors.

	\section{Methods}
	
	\begin{figure}[!t]
		\centering
		\includegraphics[width=1\columnwidth]{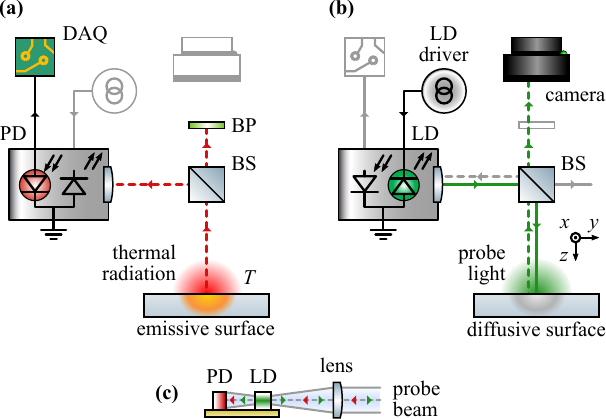}
		\caption{Simplified scheme of the hybrid sensor in the two measurement configurations, dynamically controlled by switching the current driver of the probe \glsfirst{ld} and triggering the acquisitions. (a) \gls{ld} inactive: the built-in monitor \glsfirst{pd} collects the external thermal radiation, and its photocurrent measures the surface radiance. (b) \gls{ld} emission: the scattered probe beam spot is acquired with a camera for the target distance measurement. Elements irrelevant to each specific stage are grayed out and ignored. (c) Sketch of a \gls{ld} module with embedded \gls{pd}.}
		\label{fig:scheme-on-off}
	\end{figure}
	
	The functional scheme of the hybrid sensor is illustrated in Fig.~\ref{fig:scheme-on-off}. The system core is a probe laser, integrating both a \glsfirst{ld} source and a monitor \glsfirst{pd} on the same chip. Normally, such built-in \gls{pd} is used to measure the \gls{ld} emission power, detecting the light radiated from the cavity back facet as sketched in Fig.~\ref{fig:scheme-on-off}(c). Here, the same \gls{pd} is exploited to detect external radiation, collected along the backward probe path. Therefore, when the \gls{ld} is not active as shown in Fig.~\ref{fig:scheme-on-off}(a), the \gls{pd} is used to monitor thermal radiation from an emissive surface, whose intensity depends on the surface temperature \cite{michalskiTemperature2001}, in the absence of non-thermal emission phenomena, such as gas ionization. The corresponding photocurrent is measured with a \gls{daq}. Conversely, when the \gls{ld} driver is active as shown in Fig.~\ref{fig:scheme-on-off}(b), the \gls{pd} readout is discarded, the emitted probe beam hits the diffusive target, and the light backscattered from the probe spot is detected by a camera, allowing for the determination of the sample surface morphology or distance. A \gls{bs} combines the optical beams along a common path, while a \gls{bp} spectral filter selects the probe wavelength, minimizing interferences from thermal radiation on the camera sensor. A rapid switching between the two configurations can be obtained by modulating the \gls{ld} current and synchronizing the \gls{pd} and camera acquisitions, enabling quasi-simultaneous measurements of the different parameters.
	
	The dimensional measurement relies on detecting the radiation scattered from the probe beam spot on the target surface, whose area and position depend on the beam divergence, incidence angle, and sample position. The target distance $z$ from a given reference position $z_\textrm{r}$ can be determined via triangulation by measuring the lateral position $y_\textrm{p}$ of the probe spot on the detector \cite{donadelloMonitoring2019}, following a linear relation as
	\begin{equation}
		z - z_\textrm{r} = \beta_0 y_\textrm{p}
		\label{eq:triangulation}
	\end{equation}
	where the slope coefficient $\beta_0$ defines the measurement sensitivity and depends on both the beam inclination and camera resolution.
	
	Being the embedded monitor \gls{pd} adjacent to the \gls{ld} inside the probe package, as shown in Fig.~\ref{fig:scheme-on-off}~(c), the \gls{pd} collects external radiation within the optical acceptance determined by the \gls{ld} optical path and numerical aperture \cite{wyattRadiometric2012}. Therefore, the distance and thermal radiation measurements are self-aligned on the target surface. It can be demonstrated that the \gls{pd} measurement is determined by the local spectral radiance $L_{\mathrm{e},\lambda}$, and its collection efficiency depends on the solid angle $\Omega$ defined by the probe beam divergence \cite{uedaStudies1985}. The spectral radiance, representing the radiant flux emitted at wavelength $\lambda$ by a surface at absolute temperature $T$ per unit solid angle per unit projected area per unit wavelength \cite{gaertnerOptical2012}, is given by the Planck black-body radiation law and defined as
	\begin{equation}
		L_{\mathrm{e},\lambda}(T) = \frac{2hc^2}{\lambda^5} \frac{1}{e^{hc/(\lambda k_\mathrm{B} T)}-1}
		\label{eq:radiance}
	\end{equation}
	where $h$ is the Planck constant, $c$ the light speed, and $k_\mathrm{B}$ the Boltzmann constant \cite{magunovSpectral2009}. The \gls{pd} current $I_\mathrm{e}$ is proportional to the integral of $L_{\mathrm{e},\lambda}(T)$ over its spectral bandwidth $\Delta \lambda$, over the solid angle~$\Omega$ defining the optical acceptance, and over the projected target area $A_\mathrm{p}$, therefore
	\begin{equation}
		I_\mathrm{e}(T) = \int_{\Delta \lambda} \int_\Omega  \int_{A_\mathrm{p}} R_\lambda \epsilon_\lambda(T) L_{\mathrm{e},\lambda}(T) d\lambda d\Omega dA_\mathrm{p}
		\label{eq:radiance-integral}
	\end{equation}
	where $\epsilon_\lambda(T)$ is the surface spectral emissivity, ranging from $0$ to $1$, and $R_\lambda$ is the \gls{pd} spectral response, including the optical chain transmittance. If $\Delta \lambda$ is sufficiently narrow around $\lambda$, and $\lambda T \ll hc/k_\mathrm{B}$, the \gls{pd} acts as a single-band pyrometer, and the photocurrent scales with temperature as
	\begin{equation}
		I_\mathrm{e}(T) \simeq \frac{c_1}{\lambda^5} \frac{\epsilon_\lambda(T)} {e^{c_2 / (\lambda T)}}
		\label{eq:photocurrent}
	\end{equation}
	where $c_1$ and $c_2$ are constants, assuming fixed geometry, flat response $R_\lambda$, and gray-body emissivity $\epsilon_\lambda(T)$.
	
	The probe spot defines the detection region on the target. If the probe spot is smaller than the emitting area, $I_\mathrm{e}$ does not depend on the target distance, but only on the surface radiance and probe numerical aperture \cite{nunez-cascajeroOptical2021}. However, due to probe divergence, the measurement region changes with distance: a small spot allows for the detection of local thermal effects, while a large spot averages the radiance over a wider area. Similarly, the probe spot size and beam inclination also influence the accuracy of distance measurement, since the surface morphology and diffraction may introduce distortions in the probe spot detection and affect the speckle noise \cite{dorschLaser1994}.
	
	By inverting Eq.~\eqref{eq:photocurrent}, with a single-band pyrometer the temperature can be determined only after an accurate system calibration. In practice, $\epsilon_\lambda$ depends on direction, wavelength, temperature, surface finishing, and other physical properties of the material \cite{schoppTemperature2012}, limiting this possibility. Multi-wavelength pyrometers are aimed to resolve such problems, and can be used when absolute temperature measurements are needed \cite{araujoMultispectral2017}. Nonetheless, the presented method is not intended for a precise determination of $T$. Rather, it provides a combined measurement of target distance and thermal emission, the latter being spatially averaged over the probe spot region, and serving as a proxy for the relative temperature close to a given reference condition.

	\section{Experimental Setup}
	
	\begin{figure}[!t]
		\centering
		\includegraphics[width=1\columnwidth]{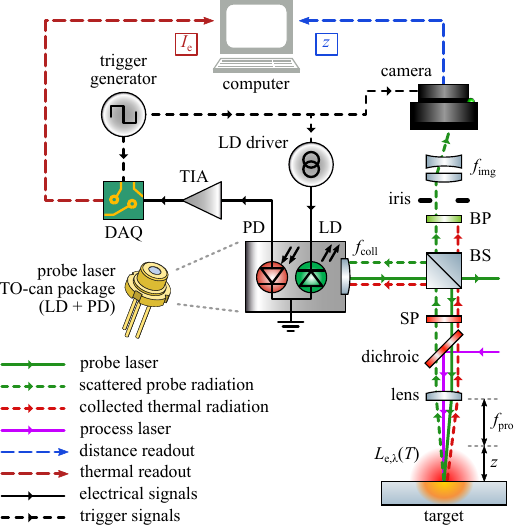}
		\caption{Experimental setup for combined thermal and dimensional measurements, implemented coaxially within a high-power laser processing system. The probe laser is housed in a package which includes both a \gls{ld} for triangulation distance measurement, and a monitor \gls{pd} for thermal emission detection. A square-wave signal modulates the \gls{ld} current and triggers the triangulation camera acquisition.}
		\label{fig:scheme-tot}
	\end{figure}
	
	Fig.~\ref{fig:scheme-tot} illustrates the complete experimental setup. The monitoring apparatus is based on a similar design demonstrated in a prior work \cite{donadelloMonitoring2019}, where a quasi-coaxial triangulation device was implemented into a laser processing machine for inline standoff distance monitoring in \gls{lmd}. In the current study, a simpler laser processing system without powder feeding is considered, operating on bulk metallic targets. The laser head can be moved with a linear-motor system. The process laser is a high-power \gls{cw} fiber laser, with wavelength $\SI{1070}{\nano\meter}$. The collimated process laser beam is deflected by a \acrlong{sp} dichroic mirror at $\ang{45}$ toward the processed sample, ensuring perpendicular incidence. A lens with focal length $f_\mathrm{pro}=\SI{200}{\milli\meter}$ focuses the beam onto the sample surface. The focused beam has waist diameter $\SI{200}{\micro\meter}$, with beam quality factor $M^2\simeq12$. It passes through a protective window and a coaxial nozzle, used for assist gas delivering. The focal plane of $f_\mathrm{pro}$ is located $\SI{10}{\milli\meter}$ outside the nozzle. The dichroic mirror also provides coaxial optical access for the monitoring setup. An additional \gls{sp} filter with $\SI{1000}{\nano\meter}$ cutoff blocks the residual scattered or reflected process radiation.

	The position $z$ is defined as the distance along the optical axis between the target and the focal plane of $f_\mathrm{pro}$. The triangulation probe is based on a \gls{cw} green diode laser, emitting $\SI{50}{\milli\watt}$ at $\SI{160}{\milli\ampere}$, with wavelength $\SI{520}{\nano\meter}$. It is collimated by an aspheric lens, with focal length $f_\mathrm{coll} = \SI{11}{\milli\meter}$ and numerical aperture $\mathrm{NA}=\num{0.26}$. The collimated beam is directed by a $50$:$50$ \gls{bs} towards the sample, transmitted by the dichroic mirror, and coaxially superimposed onto the process laser beam. A small offset of about $\SI{6}{\milli\meter}$ relative to the optical axis of $f_\mathrm{pro}$ introduces a deflection of $\ang{1.7}$, exploited for the triangulation measurement. The focused probe beam has effective waist diameter $\SI{50}{\micro\meter}$, Rayleigh range $\SI{3}{\milli\meter}$, and its divergence of $\SI{14}{\milli\radian}$ determines the sensor acceptance over the solid angle $\Omega \simeq \SI{1.6e-4}{\steradian}$. The probe beam intersects the process beam on the target surface, and it gets backscattered.
	
	The probe spot position on the sample surface is acquired by a CMOS monochrome camera. The acquisition area is $\num{96}\times\num{300}$ pixels, with pixel size $\SI{4.8}{\micro\meter}$. The camera is triggered externally, with exposure time $\SI{1.5}{\milli\second}$, and each frame is transferred to a control computer in real-time. A $\SI{1}{\milli\meter}$ diameter iris serves as spatial filter. A \gls{bp} spectral filter, with center wavelength $\SI{520}{\nano\meter}$ and full width at half maximum $\mathrm{FWHM}=\SI{10}{\nano\meter}$, prevents most of thermal radiation from reaching the camera sensor. The imaging telescope $f_\mathrm{img}$ combines a bi-concave eyepiece and a plano-convex lens, having focal lengths $\SI{-25}{\milli\meter}$ and $\SI{100}{\milli\meter}$, respectively; this, in combination with the process lens $f_\mathrm{pro}$, gives an image magnification of $\num{0.65}$. The triangulation system is calibrated as described elsewhere \cite{donadelloCoaxial2018}, resulting in a sensitivity coefficient $\beta_0=\SI{-0.106 \pm 0.003}{\milli\meter\per\pixel}$ and a reference $z_\textrm{r}=\SI{19.8 \pm 0.1}{\milli\meter}$, determined through linear fitting as defined in Eq.~\eqref{eq:triangulation}.
	
	The thermal radiance detection is performed by the silicon-based monitor \gls{pd} embedded within the TO-can package of the probe laser. The dichroic mirror defines the measurement spectral bandwidth $\Delta \lambda$ between $\SI{470}{\nano\meter}$ and $\SI{810}{\nano\meter}$, with an overall transmittance across the laser head optical elements of about $\SI{50}{\percent}$. Considering this wavelength interval and the typical experimental temperatures around $\num{1500}-\SI{2000}{\degreeCelsius}$, the approximations of Eq.~\eqref{eq:photocurrent} for a single-band pyrometer can be assumed valid. The \gls{pd} operates in photovoltaic mode. Its photocurrent $I_\mathrm{e}$ is converted into voltage by a \gls{tia}, with switchable transimpedance between $\SI{1}{\mega\ohm}$ and $\SI{8}{\mega\ohm}$. The signal is acquired by a custom \gls{daq}, comprising a pre-amplifier, a $\SI{10}{\bit}$ \gls{adc}, and a microcontroller. The acquisition operates at sampling rate $\SI{20}{\kilo\hertz}$, streaming in real-time to the control computer via serial communication.
	
	\begin{figure}[!t]
		\centering
		\includegraphics[width=1\columnwidth]{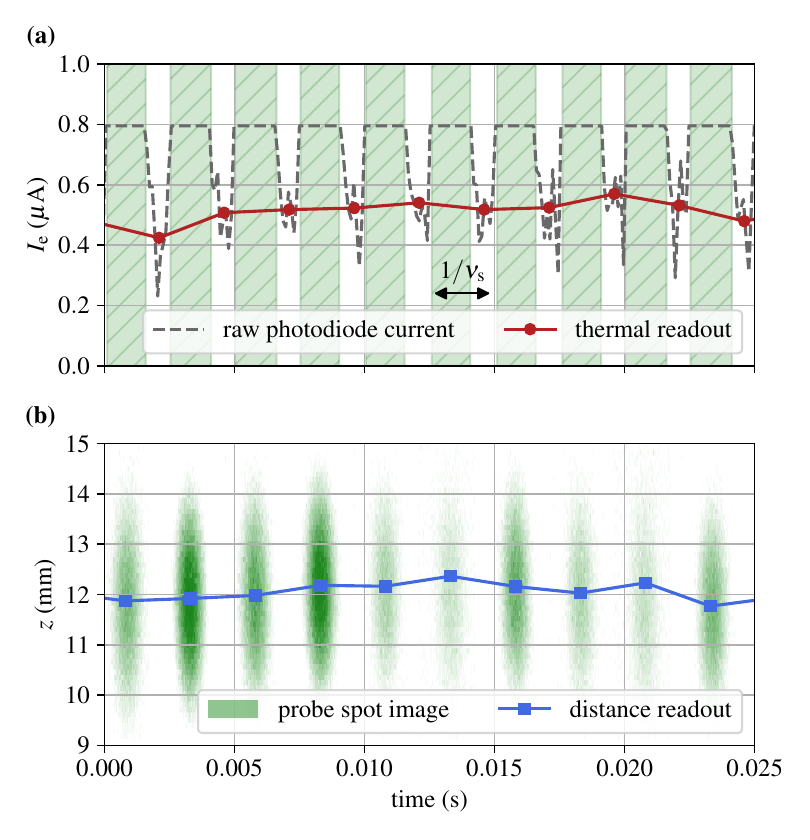}
		\caption{Combined thermal and dimensional acquisitions at \SI{400}{\hertz}. (a) Photocurrent related to thermal radiation of a laser-processed workpiece; an average sample is calculated for each interval where the \gls{ld} is not active; the shaded intervals correspond to probe laser emission and saturated \gls{pd}. (b) False-color triangulation camera frames acquired for each probe pulse; the target distance is determined from the probe spot centroid.}
		\label{fig:signals}
	\end{figure}
	
	A signal generator is used to produce a square wave at frequency $\nu_\mathrm{s}=\SI{400}{\hertz}$ and duty cycle $\SI{60}{\percent}$, used as external modulation for the \gls{ld} current driver, and to trigger the camera synchronously with each probe pulse. Fig.~\ref{fig:signals} reports an example time series, acquired while heating a stainless steel target plate with laser power $P_0=\SI{600}{\watt}$ and nominal workpiece distance $z_0=\SI{13}{\milli\meter}$. Fig.~\ref{fig:signals}(a) shows the measured photocurrent. Only values from intervals where the \gls{ld} is not emitting contribute to the average thermal readout $I_\mathrm{e}$. In contrast, during probe emission, the \gls{pd} current is ignored, being dominated by the probe light and saturating at $\SI{0.8}{\micro\ampere}$. Fig.~\ref{fig:signals}(b) shows the corresponding probe spot images acquired with the camera. The distance $z$ is determined by efficiently extracting the centroid position from the image moments, representing statistical features of the pixel intensity distribution. Temporary changes in surface reflectivity can affect the triangulation detection, leading to occasional spikes or data losses. These are mitigated by applying appropriate filters on the image parameters. The continuous trigger status toggling results in alternating measurements of $I_\mathrm{e}$ and $z$ at frequency $\nu_\mathrm{s}$, which are preprocessed and stored in real-time with a Python software.

	\section{Results and Discussion}
	
	The sensor is demonstrated in a proof-of-concept experiment, resembling a typical laser welding operation, to assess the independence between the combined measurements under realistic and dynamical conditions. A $\SI{4}{\milli\meter}$ thick AISI-304 stainless steel plate is used as primary target, with its surface positioned at $z_0=\SI{13}{\milli\meter}$. At this initial standoff distance of $\SI{23}{\milli\meter}$ the process laser beam has a spot diameter of $\SI{1.1}{\milli\meter}$. After positioning, the laser head stops for $\SI{0.5}{\second}$ without laser emission. Nitrogen gas is flowed through the $\SI{3}{\milli\meter}$ diameter nozzle at a pressure of $\SI{0.5}{\bar}$. This small inert gas flow is necessary to protect the laser head optics from material spattering and plumes while allowing a melt pool to form, having a negligible impact on the process itself, except for minor mechanical influences on the molten phase. The process emission begins at time $\SI{0}{\second}$ with power $P_0=\SI{600}{\watt}$. After a stationary interval of $\SI{0.5}{\second}$, the laser head starts a linear translation in the transverse plane at constant speed $\SI{4}{\milli\meter\per\second}$. To simulate a controlled variation in the standoff distance between the laser head and the workpiece during the transverse movement, a $\SI{1}{\milli\meter}$ step is obtained by positioning a secondary stainless steel sheet approximately $\SI{14}{\milli\meter}$ away in the translation direction.
	
	\begin{figure}[!t]
		\centering
		\includegraphics[width=1\columnwidth]{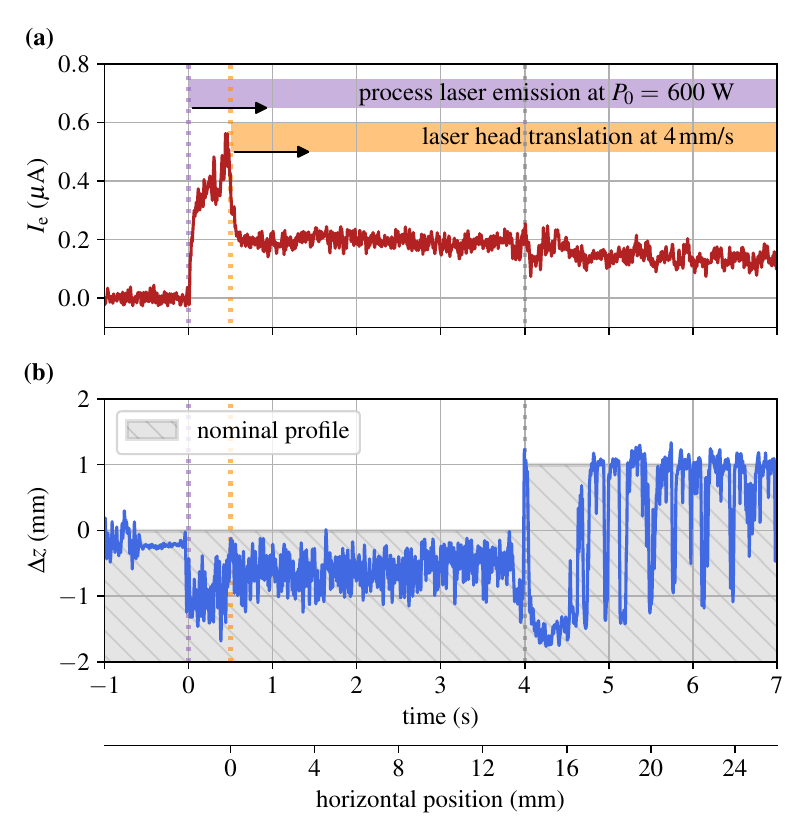}
		\caption{Combined measurements of thermal emission (a) and distance variation (b) while processing a stainless steel plate, forming a melt pool track. The process laser emission begins at time $\SI{0}{\second}$, the linear translation starts after $\SI{0.5}{\second}$. A $\SI{1}{\milli\meter}$ high step is placed along the movement path. The shaded region represents the sample nominal profile.}
		\label{fig:translation}
	\end{figure}
	
	Fig.~\ref{fig:translation} shows the radiance (a) and distance (b) measurements, smoothed using a first-order low-pass filter with a cutoff of $\SI{50}{\hertz}$. Initially, when the process starts and the laser head is stationary, the thermal emission $I_\mathrm{e}$ increases rapidly to about $\SI{0.5}{\micro\ampere}$ due to the rising temperature and melt pool formation. Simultaneously, a negative distance variation $\Delta z = z-z_0$ indicates a depression of approximately $\SI{0.5}{\milli\meter}$ below the surface. As the translation begins, the process reaches thermal equilibrium, stabilizing at a lower emission value of $\SI{0.2}{\micro\ampere}$, as a result of the reduced energy density along the melt track \cite{xiaoMonitoring2020}.

	The measurement uncertainty is estimated using the standard deviation of the signal within the filter bandwidth, calculated in the stable condition between $\SI{1}{\second}$ and $\SI{3}{\second}$. For radiance, the standard deviation of $\SI{0.02}{\micro\ampere}$ is primarily determined by noise from the photodiode and electronic circuits. For distance, the standard deviation of $\SI{0.2}{\milli\meter}$ is attributed to the sensitivity coefficient $\beta_0$ and its intrinsic calibration uncertainty, as well as to speckle noise and probe spot centroid variance. However, additional factors may also contribute to these uncertainties. First, the probe spot size, approximately $\SI{50}{\micro\meter}$, is much smaller than the processing region of about $\SI{1}{\milli\meter}$, which may cause the radiance measurement to capture local temperature fluctuations and gradients. Second, the distance measurement could be affected also by physical factors, such as laser head vibrations or melt pool ripples induced by the assist gas flow.
	
	After about $\SI{4}{\second}$ the distance probe detects the $\SI{1}{\milli\meter}$ step corresponding to the secondary plate boundary. Subsequently, $\Delta z$ exhibits rapid fluctuations exceeding $\SI{1}{\milli\meter}$ on timescales of $\SI{0.1}{\milli\second}$, likely caused by partial penetration down to the lower target, melt pool oscillations, material spattering or collapsing, instabilities during keyhole formation, and multiple probe reflections \cite{wuHigh2023,leeMechanism2002,wangUnderstanding2023}. Moreover signal loss can occasionally occur in the presence of specular reflection during such unstable conditions. The corresponding small decrease in the thermal emission indicates a reduction of the surface temperature, due to the modified laser-material interaction at the varied working distance.
	
	\begin{figure}[!t]
		\centering
		\includegraphics[width=1\columnwidth]{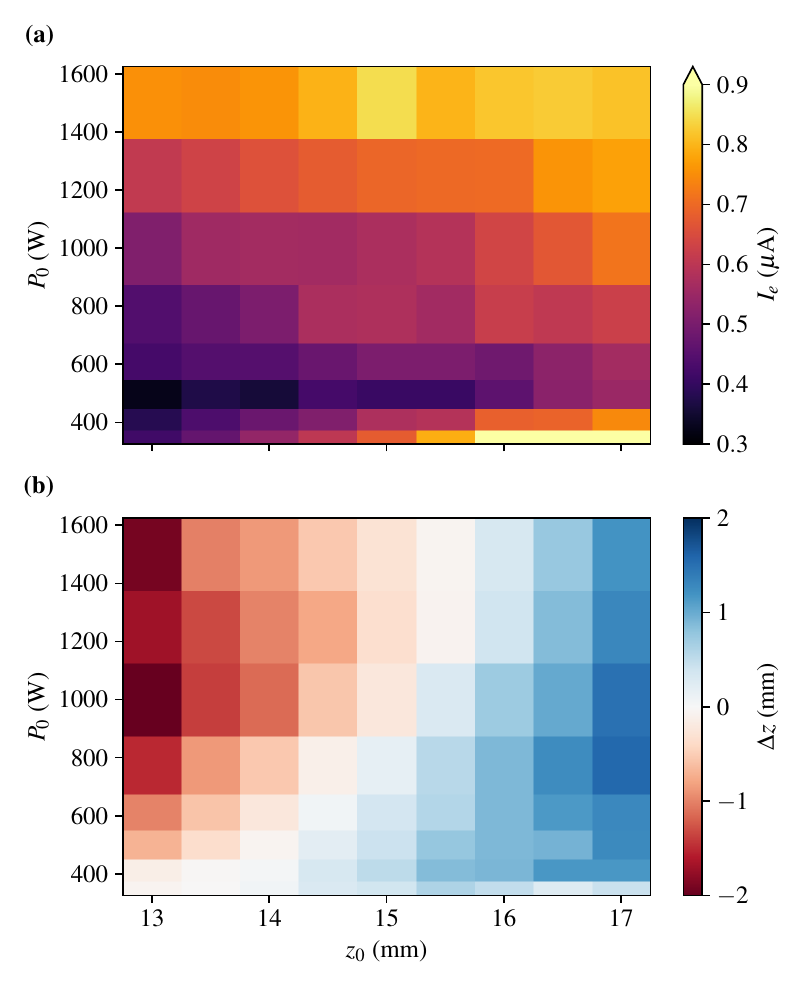}
		\caption{Color maps of average thermal emission (a) and distance variation (b) measured during stationary laser processing of stainless steel samples, at different laser powers $P_0$ and nominal distances $z_0$.}
		\label{fig:calibr}
	\end{figure}
	
	To evaluate the sensor operation in different process conditions, its response is characterized as a function of input energy and distance on a stainless steel target. The laser power $P_0$ is ranged from $\SI{350}{\watt}$ to $\SI{1.5}{\kilo\watt}$, and the nominal distance $z_0$ from $\SI{13}{\milli\meter}$ to $\SI{17}{\milli\meter}$. For each combination, the average radiance $I_\mathrm{e}$ and distance variation $\Delta z$ are measured during a laser emission interval of $\SI{1.5}{\second}$ while keeping the laser head stationary, allowing to reach thermal equilibrium due to balance between absorbed laser power and heat losses. The results are reported as color maps in Fig.~\ref{fig:calibr}. The thermal emission measurement (a) shows that, above $\SI{500}{\watt}$, the radiance increases with $P_0$, as expected from the temperature increase given by the rising optical power density on the surface. The workpiece distance has a minor influence on $I_e$, mostly referable to the varying probe spot size and position relative to local thermal gradients on the processed region \cite{nunez-cascajeroOptical2021,frunzeSystematic2013}.
	
	The measurements reported in Fig.~\ref{fig:calibr}~(a) also reveal a key physical phenomenon occurring during laser processing. Below $\SI{500}{\watt}$, especially at higher values of $z_0$, the thermal emission abruptly rises from $I_\mathrm{e}\simeq\SI{0.5}{\micro\ampere}$ to $\SI{1.2}{\micro\ampere}$, exceeding the color map scale. This apparently anomalous behavior can be attributed to strong surface oxidation observable in samples obtained under these conditions, which exhibit only partial and superficial melting surrounded by large heat-affected zones. Indeed, just before reaching the melting temperature of $\SI{1450}{\degreeCelsius}$, localized heat accumulation can occur due to high laser absorption, causing a rapid temperature rise in the surface layer, and leading to a higher apparent temperature compared to the bulk material. Above the melting point, the heat conduction mechanisms change, resulting in a more efficient dissipation, a reduced surface temperature, and the formation of a stable melt pool. These variations in thermal emission, caused by changes in material emissivity, are often observed while melting metals \cite{felicePyrometry2013,mullerTemperature2012,doubenskaiaDefinition2013}. According to Eq.~\eqref{eq:photocurrent}, the brightness temperature measured with a single-color pyrometer strongly depends on the emissivity. In particular, oxidized stainless steel exhibits a significantly higher emissivity of $\epsilon_\lambda\gtrsim\num{0.7}$ compared to its molten state, where $\epsilon_\lambda\lesssim\num{0.3}$ \cite{millsRecommended2002,balat-pichelinSpectral2022,fukuyamaNormal2022,hunnewellTotal2017}. The ratio between these values of $\epsilon_\lambda$ is of the same order as the ratio of $I_\mathrm{e}$ measured below and above the threshold power of approximately $\SI{500}{\watt}$, thus explaining the observed increase in thermal emission at low power densities in terms of oxidation effects. Similar changes in the material brightness have been reported in other laser processing studies as well, near the melting point or during the transition between conduction and keyhole regimes \cite{dasilvaThermal2022,laneTransient2020,allenEnergyCoupling2020}.
	
	The complementary characterization of the dimensional measurement is reported in Fig.~\ref{fig:calibr}~(b), showing the average distance variation for each condition. At low powers, where a melt pool is not fully formed, the measured distance matches the nominal value. Above the melting threshold, $\Delta z = z-z_0$ depends strongly on the initial distance $z_0$, with weak influence from laser power. This suggests that the melt pool depth is primarily determined by the position of the process laser focal plane. A closer laser head leads to a depression in the melt pool, likely due to the smaller spot size and stronger coaxial assist gas flow, expelling the molten material. For larger distance values, porosity, spatters and droplets can cause surface swelling \cite{gaoObservation2017}, exceeding $\SI{1}{\milli\meter}$. However, optical distortions in the absolute triangulation measurement cannot be completely excluded \cite{dorschLaser1994}. These might be caused by speckle noise \cite{csencsicsReducing2023}, multiple reflections within the keyhole \cite{flemingSynchrotron2023}, or parallax errors arising from the probe beam offset relative to the optical axis, especially when probing irregular shapes.

	\section{Conclusion}
	
	This work presented an innovative and versatile technique for combined thermal and dimensional measurements. The working principle is based on a simple and cost-effective triangulation device, implemented in a quasi-coaxial configuration into a laser processing setup. The thermal radiance measurement is performed without requiring additional components, since the monitor photodiode normally embedded in laser diode packages is exploited as a single-band pyrometer by continuously switching the probe laser emission. Indeed, a single instrument can conveniently replace other multi-sensor approaches. The detection is self-aligning, and performed over a localized area determined by the probe beam spot. The hybrid sensor was successfully demonstrated for high-speed monitoring of thermal emission and distance of a melt track generated by a high-power laser on a stainless steel target. The synchronized measurements allowed to identify the melt pool formation and the transition between different regimes, like metal oxidation and melting, providing insights into process stability and dynamics.
	
	Thanks to its low intrusiveness, the system can be easily integrated for coaxial in-line monitoring in various laser processes, including penetration welding, cutting, ablation, and additive manufacturing techniques, such as \gls{lmd} or \gls{slm}. While absolute temperature measurements would require specific calibrations, which fall outside the scope of this study, the method is particularly suitable for relative measurements, offering valuable applications in process stabilization and optimization. The preliminary results showed that the hybrid sensor can provide real-time simultaneous readings of distance and radiance at rates of hundreds of \si{\hertz}. In the future, these data streams could be used to feed closed-loop control systems, allowing them to actively adjust process parameters such as laser power, standoff distance, and translation speed, to stabilize the error signal relative to the desired thermal and dimensional setpoint. Moreover, it can be used for both online monitoring of defect formation during the process, and offline quality check through post-process optical profilometry during the workpiece cooling down.
	
	The sensor design could be improved by substituting the camera, which may be unnecessary for one-dimensional distance measurements, such as in triangulation. Instead, a photodiode array or an optical position-sensitive detector could reduce the setup complexity and costs, improve the readout speed, and eliminate the need for a control computer, since such sensors can be easily read with \glspl{adc} and preprocessed by embedded systems like microcontrollers or \acrshortpl{fpga}. Incorporating multiple probes with photodiodes sensitive to different wavelengths would enable more precise temperature measurements, following the principle of multi-channel pyrometry. Moreover, the methodology can be adapted from triangulation to \gls{lidar}, reflectometry, interferometry, \gls{oct}, fiber sensing, or any other optical detection technique where the probe embeds a detector that can be exploited in a bidirectional way. In perspective, this dual-mode integrated optical technique can offer valuable tools for researchers and engineers in science, technology, and industry, with applications in diverse fields, from remote monitoring of high-temperature processes to biological and environmental sensing.

	\section*{Acknowledgment}
	The author would like to sincerely thank Barbara Previtali and Ali Gökhan Demir from the Politecnico di Milano for providing the inspiring scientific context that contributed to the development of this work. 
	Adige SpA of the BLM Group, in particular, Maurizio Sbetti and Daniele Colombo, are acknowledged for their valuable support of the activity. 
	Giorgio Brida, Valentina Furlan, Valentina Finazzi and Marco Scapinello are thanked for the fruitful discussions and proofreading. 
	
	Adige SpA is the current assignee of the following granted patent, inherent to the methodology presented in the paper: S. Donadello, and B. Previtali, ``Combined Optical System for Dimensional and Thermal Measurements, and Operating Method Thereof'', European Patent Office EP3889539B1, 2023. 
	The work presented in the paper has been partially funded with the contribution of the Autonomous Province of Trento, Italy, through the Regional Law 6/99 (Project LT 4.0).

	\bibliographystyle{IEEEtran}
	\bibliography{bibliography}
	
		
	
\end{document}